# A Machine Learning Nowcasting Method based on Real-time Reanalysis Data

Lei Han[1,2], Juanzhen Sun[3], Wei Zhang[1,2], Yuanyuan Xiu[1], Hailei Feng[1], and Yinjing Lin[4]

[1]College of Information Science and Engineering, Ocean University of China, Qingdao, Shandong, China

[2]Institute of Urban Meteorology, China Meteorological Administration, Beijing, China

[3]National Center for Atmospheric Research, Boulder, CO

[4]National Meteorological Center, China Meteorological Administration, Beijing, China

Corresponding author: Juanzhen Sun (sunj@ucar.edu), Wei Zhang (weizhang@ouc.edu.cn)

**Abstract**

Despite marked progress over the past several decades, convective storm nowcasting remains a challenge because most nowcasting systems are based on linear extrapolation of radar reflectivity without much consideration for other meteorological fields. The variational Doppler radar analysis system (VDRAS) is an advanced convective-scale analysis system capable of providing analysis of 3-D wind, temperature, and humidity by assimilating Doppler radar observations. Although potentially useful, it is still an open question as to how to use these fields to improve nowcasting. In this study, we present results from our first attempt at developing a Support Vector Machine (SVM) Box-based nOWcasting (SBOW) method under the machine learning framework using VDRAS analysis data. The key design points of SBOW are as follows: 1) The study domain is divided into many position-fixed small boxes and the nowcasting problem is transformed into one question, i.e., will a radar echo > 35 dBZ appear in a box in 30 minutes? 2) Box-based temporal and spatial features, which include time trends and surrounding environmental information, are constructed; and 3) The box-based constructed features are used to first train the SVM classifier, and then the trained classifier is used to make predictions. Compared with complicated and expensive expert systems, the above design of SBOW allows the system to be small, compact, straightforward, and easy to maintain and expand at low cost. The experimental results show that, although no complicated tracking algorithm is used, SBOW can predict the storm movement trend and storm growth with reasonable skill.

## 1 Introduction

Although there has been much progress over the past several decades, very short-term convective storm forecasting, or "nowcasting," remains challenging. Existing nowcasting methods can be classified into three categories: extrapolation techniques based on radar data, numerical weather prediction (NWP) models, and knowledge-based expert systems that blend NWP and extrapolation techniques [*Wilson et al.,* 1998, 2010; *Sun et al.,* 2014].

Extrapolation techniques include cross-correlation tracking [*Rinehart and Garvey,* 1979; *Tuttle and Foote*, 1990; *Li et al.,* 1995; *Lai,* 1999] and centroid tracking [*Austin and Bellon,* 1982; *Rosenfeld,* 1987; *Handwerker*, 2002; *Han et al.*, 2009] techniques. Centroid tracking can be used to obtain various properties of a single storm cell, such as storm area, volume, top, base, etc. Storm Cell Identification and Tracking [*SCIT; Johnson et al.,* 1998] and Thunderstorm Identification, Tracking, and Nowcasting [*TITAN; Dixon and Wiener,* 1993] are two well-known centroid-type nowcasting algorithms. In contrast, the cross-correlation tracking method does not aim at single storm cells, but can provide the motion vectors for all radar echoes. Although widely used, none of these extrapolation techniques are able to forecast convective storm initiation, and they have no or only limited ability in forecasting storm growth and decay, thus limiting their nowcasting accuracy [*Dixon and Wiener*, 1993; *Wilson et al.*, 2010]. Nowcasting techniques were first developed with radar observations; recently, attempts have been made to incorporate satellite data using similar techniques as well as passive remote sensing-based techniques, such as infrared (IR) temperature time-differencing and multispectral IR channel differencing techniques [*Mecikalski and Bedka,* 2006; *Sieglaff et al.,* 2011, 2013].

Although there has been a great deal of progress in the nowcasting application of high-resolution and convection-permitting NWP, it is still far from being adequate for this purpose [*Weisman et al.*, 2008; *Sun et al.*, 2014]. Many problems remain to be addressed, such as the need for running NWP with model resolutions of less than a few kilometers, dealing with the spinup problem, rapid model error growth at the convective scale, etc.

The use of expert systems that attempt to blend the strengths of extrapolation techniques and NWP, such as Auto-Nowcaster[*ANC; Mueller et al.,* 2003], Nowcasting and Initialization of Modelling Using Regional Observation Data System [*NIMROD; Golding,* 1998], and Short-Range Warning of Intense Rainstorms in Localized Systems [*SWIRL; Yeung et al*., 2009], is becoming increasingly common [*Wilson et al*., 2010]. However, other than being complicated and requiring input from many data sources as well as large maintenance efforts, these expert systems depend on the forecast accuracy of NWP models and suffer the same problems as the direct use of NWP for nowcasting.

The variational Doppler radar analysis system (VDRAS) is a high-resolution data assimilation system that was designed to retrieve unobserved meteorological variables of wind, temperature, and humidity at the convective scale, with frequent updates at intervals of less than 20 minutes, by assimilating radial velocity and reflectivity from single or multiple Doppler radars [*Sun and Crook*, 1997, 2001; *Sun et al.*, 2010). Because the advanced four-dimensional variational (4DVAR) data assimilation technique is used with a cloud-model (which is its constraint over a short assimilation window), VDRAS is able to produce frequently updated analysis in a dynamically consistent manner. In addition to radar data, VDRAS can also assimilate data from in-situ observations, such as radiosondes, profilers, surface networks, VAD analysis, and mesoscale model analysis. Over the past several decades, VDRAS has been run in many weather service offices throughout the world and has proven to be an effective tool for providing useful real-time information to nowcast convective weather. As VDRAS analysis are highly dependent on high-resolution radar observations, they have been shown to be quite accurate in convective situations [*Sun et al.* 2010]. The retrieved meteorological fields have been used directly by forecasters, and as input into expert systems to nowcast storm initiation and location [*Mueller et al.*, 2003].

In this paper, we describe a new method of using VDRAS analysis for thunderstorm nowcasting. The method employs VDRAS analysis data to build a Support vector machine Box-based nowcasting (SBOW) algorithm under the machine-learning framework. SBOW constructs box-based temporal and spatial features, which includes time trends and surrounding environmental information. These box-based constructed features are used to train the support vector machine (SVM) classifier and then the trained classifier is used to make predictions. The concise design of SBOW makes the system small, compact, straightforward, and easy to maintain and expand at low cost. Only VDRAS and radar data are needed. In this study, we used VDRAS analysis of five heavy rainfall/flash flood cases that occurred in eastern Colorado to train the system; two similar cases were then used for the predictions. The nowcasting skill of SBOW was verified against radar observations. The experimental results showed that, although we used no complicated tracking algorithm, SBOW could predict storm movement trends and storm growth with reasonable skill. Although in the current study SBOW was applied to VDRAS analysis data to show its potential, the method can be applied to any other meteorological analysis datasets.

This paper is organized as follows. Section 2 describes the data used in this study. Section 3 introduces the methodology, and Section 4 describes the analysis results of case studies. Finally, the conclusions are presented in Section 5.

**2 Data**

The data used in this study includes reflectivity data from the KFTG WSR-88D radar located in Denver, USA and VDRAS analysis data. The KFTG reflectivity data were also used to verify the nowcasting results. For the convenience of data processing, the radar reflectivity data were transformed from spherical coordinates into Cartesian coordinates with a horizontal spatial resolution $0.01° \times 0.01°$ (about 1 km × 1 km) and a vertical resolution 1 km.

The grid resolution of the VDRAS data used in this study was 3 km in the horizontal direction and 300 m in the vertical direction with a grid mesh of $280 \times 230 \times 20$. The interval of two successive VDRAS outputs is 15 minutes. For consistency with the radar data, all VDRAS data were interpolated into the same grid as radar data with a horizontal spatial resolution of $0.01°$.

The radar and VDRAS data of seven historic heavy rainfall events in the Colorado front range area of the Rocky Mountains used in this study (8 – 9 August 2008, 28 – 29 July 2010, 9 – 10 August 2010, 13 – 14 July 2011, 14 – 15 July 2011, 6 – 7 June 2012, and 7 – 8 July 2012) were collected from a retrospective study of historical heavy rain/flash flood cases conducted by the Short Term Explicit Prediction (STEP) Program of NCAR. STEP is a multi-NCAR laboratory activity aimed to improve the short-term forecasting of high-impact weather, such as severe thunderstorms, winter storms, and hurricanes (http://www.rap.ucar.edu/projects/step/).

Figure 1 shows the VDRAS analysis domain and our study domain. The smaller study domain was chosen to eliminate near-boundary areas and high mountain regions where observations were scarce. The VDRAS analysis data includes basic meteorological fields with three velocity components: temperature, humidity, and microphysics, and derived fields such as divergence.

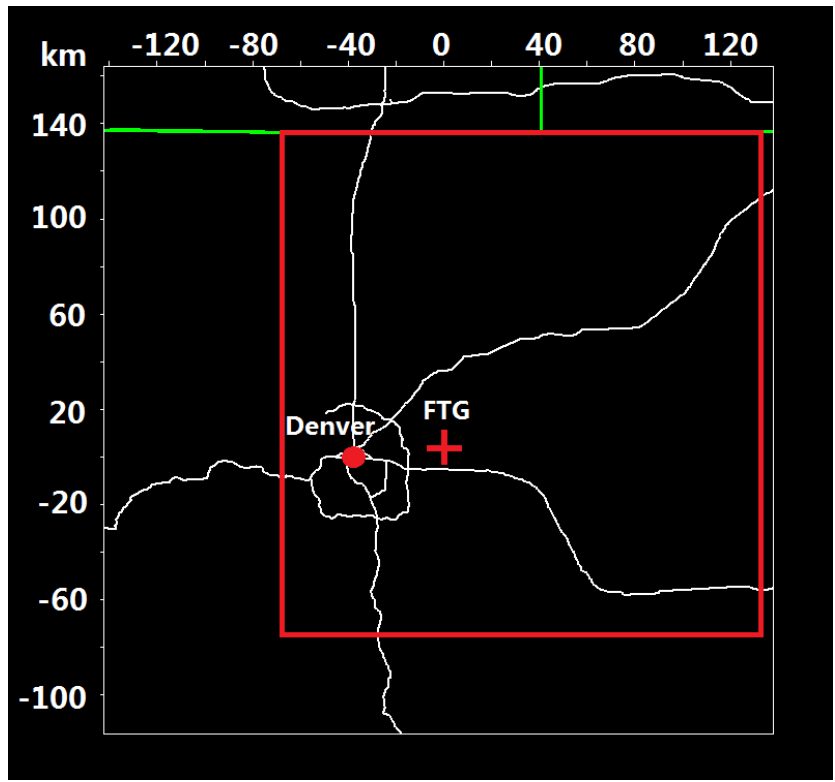

**Figure1.** The VDRAS analysis domain and the study domain (red rectangle). The location of the KFTG radar near Denver is shown by the red cross. The green lines indicate the state borders and the white lines are the major highways.

## 3 Methodology

The key design features of the SBOW algorithm are as follows: 1) the whole method is based on small boxes. The study domain is divided into many position-fixed small boxes, and the nowcasting problem is transformed into one question, i.e., will a radar echo > 35 dBZ appear in a box in 30 minutes? 2) box-based temporal and spatial features, which include time trends and surrounding environmental information, are constructed and 3) a machine learning framework is applied to perform the nowcasting task, i.e., the box-based constructed features are used to train the SVM classifier first, and the trained classifier is then used to make predictions. It should be noted that in this study we divide all the data into three independent subsets for training, cross validation, and testing, respectively.

Figure 2 presents an overview of the flow of the algorithm. The overall method consists of three main components: 1) box-based temporal feature construction; 2) box-based spatial feature construction; and 3) application of SVM, a powerful machine learning method, to train the classifier and make 30-min forecasts. These components are described below.

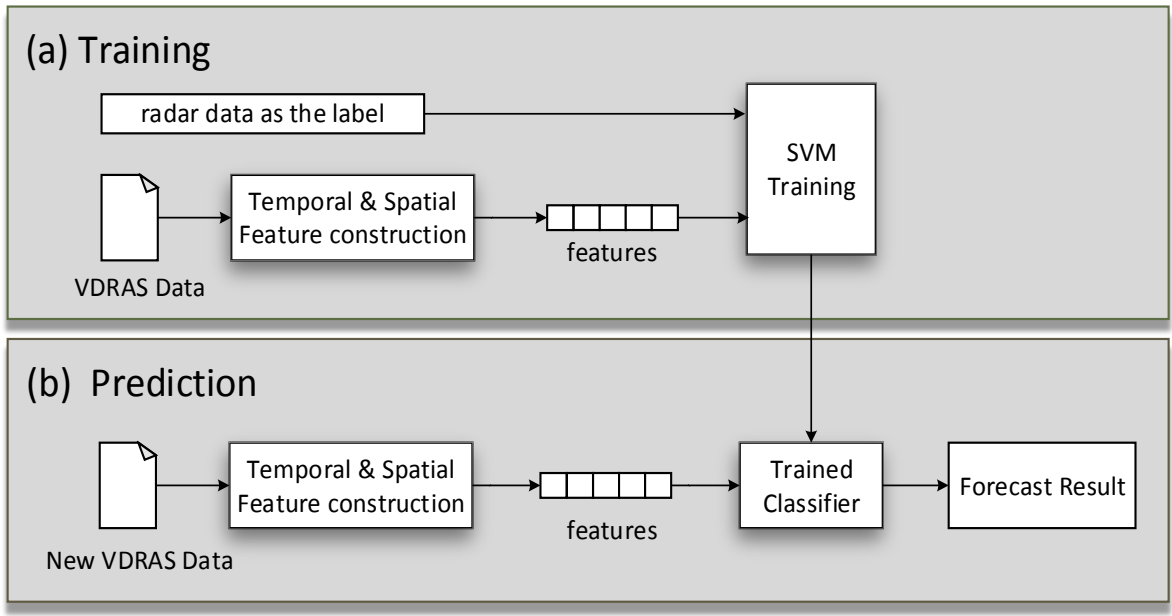

**Figure2.** Flowchart of the SBOW algorithm.

3.1 Box-based temporal feature construction

The first step in developing the SBOW algorithm was to select candidate features or predictors that could be used for training. Among the analysis variables of VDRAS, we first chose vertical wind *(w)* and the perturbation temperature (*pt*, subtracted from the horizontal mean), which is proportional to the buoyancy. We then constructed the temporal features of their time trends, i.e., *dw* and *dpt*, respectively. The four variables were used in the SBOW algorithm as the candidate predictors.

Vertical velocity is closely related to the low-level convergence and plays an important role in storm initiation and development [*Wilson and Mueller*, 1993]. A large value of w indicates strong lifting, which is one of the necessary ingredients to achieve deep convection [*Doswell*, 1992]. Buoyancy represents the vertical acceleration of an air parcel resulting from an unstable atmosphere. It is the driving force and plays a key role in deep convection [*Wallace and Hobbs* 1977]. The temporal trend contains information about storm growth/decay and thus can also play an important role in convective nowcasting. [*Roberts and Rutledge*, 2003; *Mecikalski and Bedka*, 2006; *Sieglaff et al.,* 2011] have shown that satellite infrared (IR) cloud-top temperature trend information is very useful in forecasting convective initiation.

Instead of computing *w, pt, dw*, and *dpt* at each pixel, these features were computed in a 3D box (0.06° × 0.06°, 20 levels). The values of *w* and *pt* are calculated first. Particularly, *w* is the maximum value between the earth's surface and 4 km above the surface in the 3D box, and *pt* is the maximum value above 4 km in the 3D box. The choice of the vertical layers for *w* and *pt*, although being somewhat arbitrary, is based on the consideration that the lower level vertical velocity and higher-level latent heating could play greater roles in convective initiation. After obtaining *w* and *pt* in each box at each time step, *dw* and *dpt* can be determined by computing their respective differences between current and previous time steps. Therefore for each box at a time step, we can obtain four features: *w, pt, dw,* and *dpt.*

## 3.2 Box-based spatial feature construction

Weather phenomena are not only continuous in time but also in space, i.e., each box is impacted by its neighboring boxes, so it is necessary to construct spatial features to take these impacts into account.

As shown in Figure 3, each box has four features (*w, pt, dw, dpt*). However, instead of only assigning its own four features to the box located at (*i, j*), we also assigned the same features of its neighboring eight boxes. Thus, there are 4 × 9 = 36 features for the box located at (*i, j*), and the same is true for all other boxes. This allows the SBOW algorithm to consider the surrounding information around the center box, which can be important for a successful nowcast.

Although we believe that our choice of the four candidate predictors has a good physical basis, there could be other predictors that would be useful to help improve the nowcast, for example, moisture, vertical wind shear, etc. But it is well known that, in most instances, SVM classifiers are more accurate with some feature reduction. More features do not mean better classification performance. We performed a greedy feature selection, i.e. the feature set was initialized from empty set first, and then we iteratively added one feature that maximized the classification performance among the rest features. {*w, pt, dw, dpt*} was finally selected.

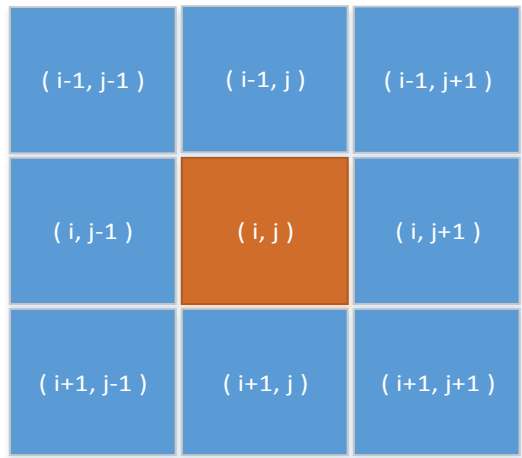

**Figure3.** Illustration of box-based spatial feature construction. The surrounding boxes used for additional spatial features are indicated by the blue color.

## 3.3 Application of SVM

SVM is a supervised learning method. Through learning from a labeled training dataset, it constructs an optimal hyperplane in a high-dimensional space for classification [*Boser et al.* 1992; *Cortes and Vapnik,* 1995; *Vapnik,* 1998]. SVM has become a mature, powerful tool and been used in atmospheric research [ *Lee et al.,* 2004; *Mercer et al.*, 2008; *Felker et al.,* 2011; *Zhuo et al.* 2014].

### 3.3.1 Description of SVM

The basic SVM is a two-class classifier. Given a training set with *n* dimensional feature vectors $x_i \in R^n$ and its corresponding label $y_i \in \{-1,1\}$, SVM solves the following optimization problem [*Chang and Lin*, 2011]:

$$\min_{w,b,\xi} \left(\frac{1}{2} w^T w + C \sum_{i=1}^{l} \xi_i\right) \qquad (1)$$

subject to $y_i(w^T \phi(x_i) + b) \geq 1 - \xi_i,$

$\xi_i \geq 0, i = 1, ..., l$

where $\phi(x_i)$ maps $x_i$ into a higher dimensional space and $C$ is the penalty parameter of the error term. $K(x_i, x_j) = \phi(x_i)^T \phi(x_j)$ is defined as the kernel function. We use the radial basis function (RBF) kernel:

$$K(x_i, x_j) = \exp(-\gamma \|x_i - x_j\|^2), \gamma > 0 \qquad (2)$$

where $\gamma$ is kernel parameter. The pair of penalty and kernel parameter, $(C, \gamma) = (4, 1.25)$, is chosen by a grid search on the 5-fold cross validation (5-CV) of the training set [*Chang and Lin*, 2011]. This grid search iteratively test classification performance on 5-CV for values of $C$ ($2^{-5}$, $2^{-4}$, $2^{-3}$,..., $2^{15}$) and RBF kernel parameter $\gamma$ ($2^{-15}$, $2^{-14}$, $2^{-13}$,..., $2^{3}$). Among commonly used kernels, besides RBF, the polynomial kernel is also an option. But for our study, the RBF kernel performs better.

It should be noted that VDRAS, being a 4DVAR method, also suffers from NWP model errors, and these errors will therefore degrade the nowcasting algorithm performance. However, a major strength of machine learning methods is that they could potentially mitigate the effects of systemic NWP model errors or observation errors. This strength sources from the generalization ability of machine learning [*Bishop* 2006; *Mohri et al*. 2012]. A core objective of a classifier is to generalize from its experience. Generalization in this context is the ability of a classifier to perform accurately on new, unseen data after constructing a model on training set whilst preventing the model from overfitting to the training set. In the view of the structural risk minimization principle of SVM, the generalization is fulfilled by the regularization term $C \sum_{i=1}^{l} \xi_i$ of the optimization function: $\min_{w,b,\xi}\left(\frac{1}{2} w^T w + C \sum_{i=1}^{l} \xi_i\right)$ [*Vapnik and Chervonenkis*, 1971; *Vapnik*, 2000].

Use of SVM consists of two steps: training and prediction. They are described below.

### 3.3.2 Training

The study domain was divided into 1,209 boxes. If there are 10 time levels of VDRAS dataset for one convective weather event, the total number of boxes will be 12,090. At each time level, each box has 36 features, including four features from the box itself, and another 32 features from its neighboring eight boxes. For training purposes, it is necessary to label each box. For a box at time t, if there is a radar echo > 35 dBZ at time t + 30 minutes, this box is labeled "1," which means "a convective storm will happen in this box in 30 minutes." Otherwise, this box will be labeled "0," which means "no convective storm will occur in this box in 30 minutes."

We used five historic heavy rainfall/flash flood events over the Front Range area for training (8 – 9 August 2008, 28 – 29 July 2010, 9 – 10 August 2010, 13 – 14 July 2011, 14 – 15 July 2011), and obtained 195,858 labeled boxes as the training and validation dataset. Each labeled box has 36 features. Each feature is scaled to [-1,1] by the min-max normalization.

After training, the obtained SVM classifier has acquired the "knowledge" to answer the question: given the 36 features of a box, will a radar echo > 35 dBZ appear in this box in 30 minutes?

### 3.3.3 Prediction

With the trained SVM classifier, when given new VDRAS data, we could use this classifier to make predictions. For example, at time t, when new VDRAS data arrive, it is divided into 1,209 boxes and the 36 features are calculated for each box. These 36 features are used as inputs into the trained SVM classifier. If the output of SVM is 1, we would predict that there will be a convective storm in this box in 30 minutes; this box will be marked as a red rectangle (examples will be given in the next section).

Given the RBF kernel and the size of feature set, the computational cost of applying the trained SVM model mainly depends on the number of support vectors. In our case, there are only 18709~19026 support vectors in 5-CV. Typically, it costs only ~0.005 seconds to make a prediction for one sample.

## 4 Experiments and Analysis

### 4.1 Comparison results of five machine learning methods

The contingency table approach (Donaldson et al. 1975) was used in this study to evaluate the nowcast results. The probability of detection (POD), false alarm ratio (FAR) and critical success index (CSI) are calculated. At each box, a success (S) occurs when this box is classified as 1 (active) and there is a radar echo greater than 35 dBZ in the forecast time in the same box (active), a failure (F) occurs when the truth box is active while the forecast box inactive, and a false alarm (A) occurs when the truth box is inactive while the forecast box active. Thus, POD = S/(S + F), FAR = A/(S + A), and CSI = S/(S + F + A) [*Dixon and Wiener* 1993].

First, we compare SBOW performance with four other machine learning methods using 5-fold cross validation (5-CV). The other four methods are: logistic regression [Cox, 1958], J48, Adaboost and Maxent. Although the last three methods are less used by atmospheric scientists, they are well known in machine learning community. J48 is an open source implementation of the C4.5 decision tree algorithm [*Quinlan*, 2014; *Frank et al.*, 2016]. Maxent is a maximum entropy modeling method aiming to find the best model $p^*$ with maximum entropy:

$$p^* = \text{argmax}_{p \in C} H(p) \tag{3}$$

where $H(p)$ is the conditional entropy of the posterior probability of the data, $C$ is the set of all possible models which satisfy specific feature statistics [*Jaynes*, 1957; *Phillips et al.*, 2004]. Adaboost is an ensemble method that combines a group of weak classifiers to construct a strong classifier [*Freund and Schapire*, 1995; *Viola and Jones*, 2004]. At each iteration $t$, a weak classifier is selected and assigned a coefficient $\alpha_t$ such that the following sum training error $E_t$ is minimized:

$$E_t = \sum_i E[F_{t-1}(x_i) + \alpha_t h(x_i)] \tag{4}$$

where $F_{t-1}(x_i)$ is the classifier that has been built up to the previous stage of training, $\alpha_t h(x)$ is the weak classifier that is being considered for addition to the final classifier.

All methods used the same training data. Each training sample has 36 features. We randomized the sample set by shuffling all samples for cross-validation. All instances were subjected to 5-CV training and tests. The means and standard deviations (σ) for CSI, POD, and FAR are shown in Table 1. SVM had the highest CSI and POD value and the lowest FAR among all the methods.

Table 1: Comparison of five machine learning methods on 5-CV

| Approach | POD(±σ) | FAR(±σ) | CSI(±σ) |
| --- | --- | --- | --- |
| SBOW (SVM) | 0.6050±0.0058 | 0.5243±0.0082 | 0.3631±0.0064 |
| Logistic Regression | 0.5111±0.0361 | 0.5799±0.0360 | 0.2980±0.0086 |
| J48 | 0.5100±0.0154 | 0.5336±0.0177 | 0.3066±0.0296 |
| Adaboost | 0.5740±0.0205 | 0.5892±0.0062 | 0.3146±0.0044 |
| Maxent* | 0.5033±0.0232 | 0.5880±0.0253 | 0.2922±0.0090 |

* Maxent is from http://homepages.inf.ed.ac.uk/lzhang10/maxent.html

4.2 Comparison results of SBOW with TITAN

4.2.1 Qualitative Analysis

Two heavy rainfall events (6 June 2012 and 7 July 2012) were used to compare the performances of SBOW and TITAN. For each of the two cases, the SBOW was run for the period between 21UTC to 00UTC, producing 12 30-min nowcasts at VDRAS analysis times that were available every 15 minutes. The radar composite reflectivity images of these two cases are shown in Figures 4 and 5, respectively, with a 30-min interval to show the evolution of the systems. In the following section, a qualitative analysis will be presented, followed by a quantitative analysis, which will together examine the performance of SBOW as compared with TITAN.

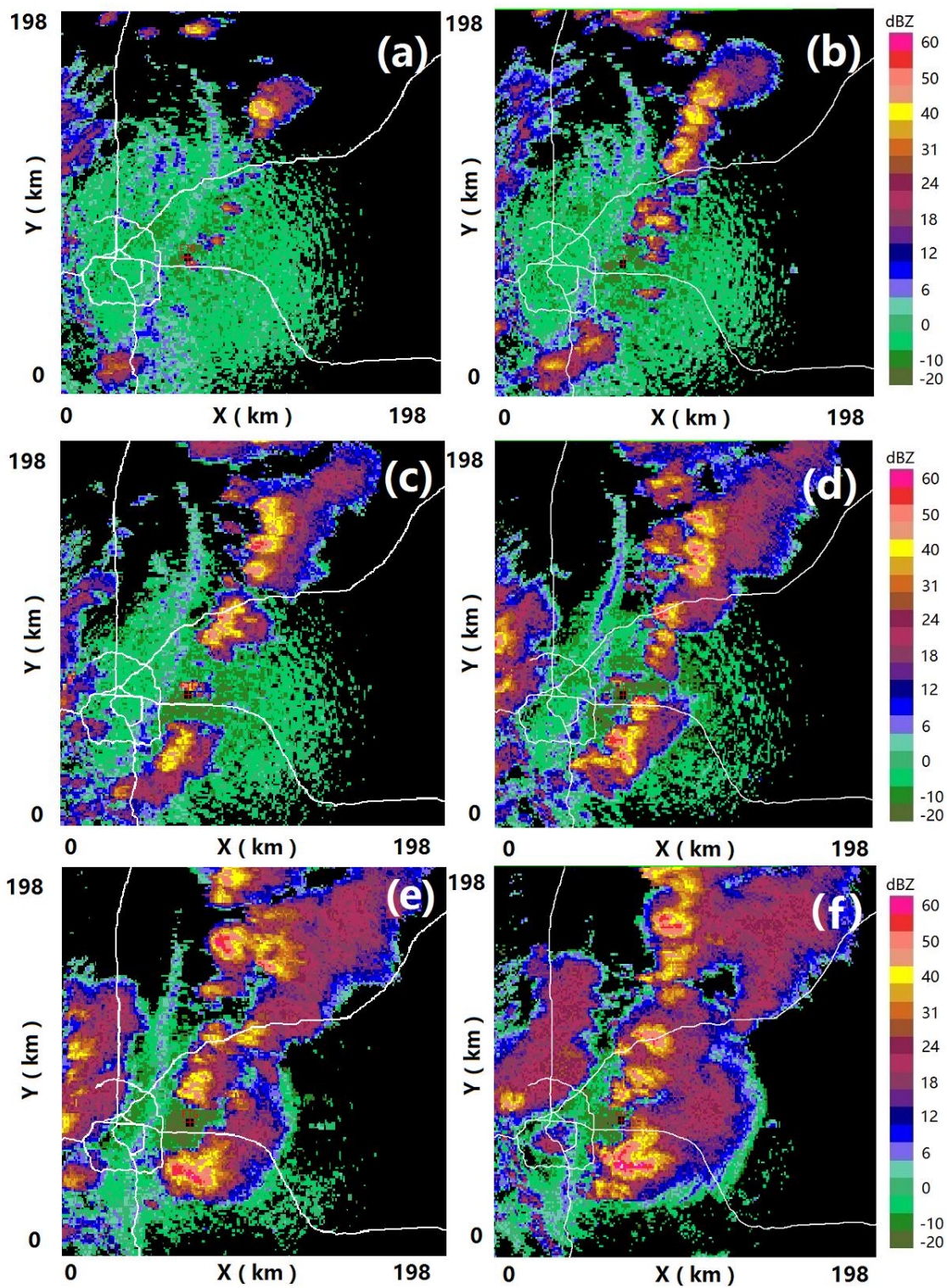

**Figure 4.** The KFTG radar composite reflectivity images at (a) 2130, (b) 2200, (c) 2230, (d) 2300, (e) 2330UTC on 6 June, and (f) 0000 UTC on 7 June 2012.

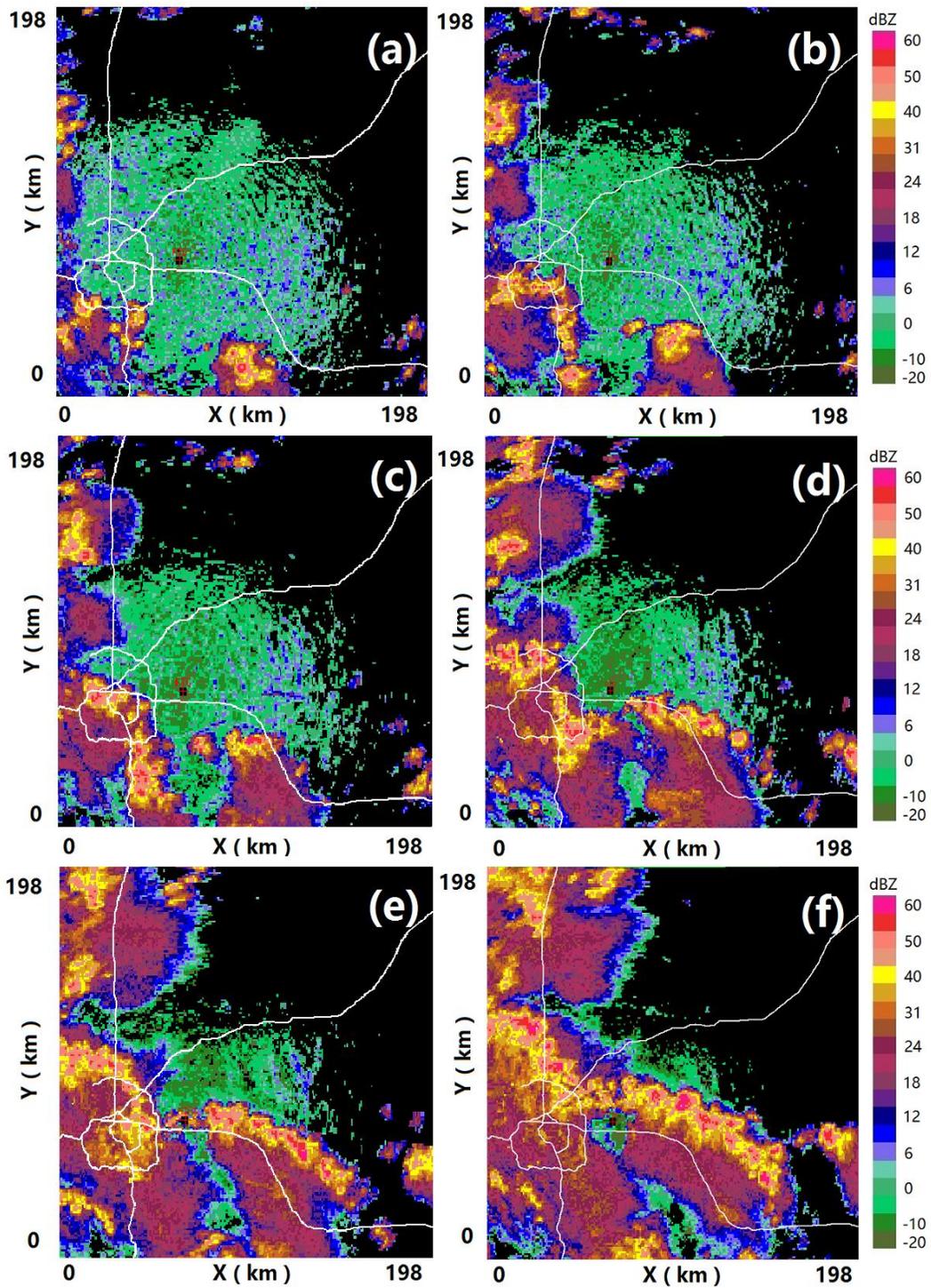

**Figure 5.** The KFTG radar composite reflectivity images at (a) 2130, (b) 2200, (c) 2230, (d) 2300, (e) 2330UTC on 7 July, and (f) 0000 UTC on 8 July 2012.

We first present the 30-min forecast results for a storm growth case of 7 July 2012 over the southeast Denver area. Figure 6 shows the observed radar reflectivity overlaid with SBOW 30-min nowcasts (red boxes) and the corresponding verifications (those correctly predicted boxes are marked as black boxes). Cyan polygons represent TITAN 30-min nowcasts. The verification shown by the black boxes in Figure 6(b) confirms that the 30-min SBOW nowcast agrees well with the radar 35+dBZ echoes. In comparison, TITAN can only extrapolate the existing storm (the bottom-left storm in Figure 6(a)), and is not able to forecast this dramatic storm growth well. This is typical shortcoming of all extrapolation methods.

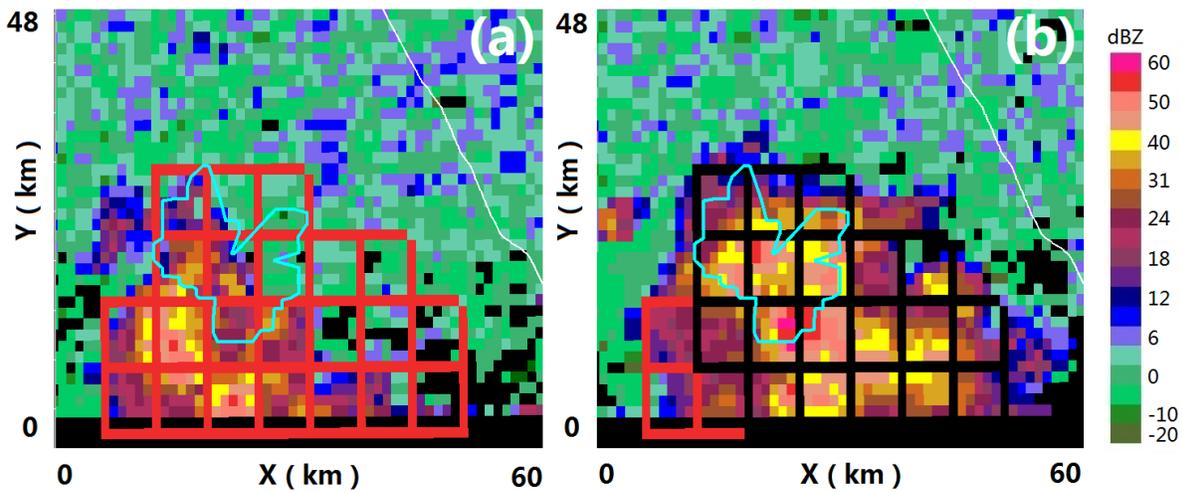

**Figure 6.** Radar composite reflectivity from KFTG; the red boxes represent the 30-min SBOW nowcasts. Cyan polygons represent TITAN 30-min nowcasts. The boxes in the right column are marked in black to represent those being correctly predicted by SBOW. (a). The SBOW and TITAN 30-min nowcasts at issue time, 2055 UTC 7 July 2012. (b). The same nowcasts superpositioned over reflectivity at verification time 2125 UTC.

The next two figures (Figures 7 and 8) show examples of convective initiation. Here we use "convective initiation" or CI to refer to a storm that is grown from "scratch" (fewer than 35 dBZ echoes) rather than from an existing storm nearby with greater than 35 dBZ echoes. The verification on the right panels in both figures shows good agreement between the SBOW nowcast and the observed reflectivity. Meanwhile, TITAN just extrapolates existing storms and misses the newly born storms in front of the old storms.

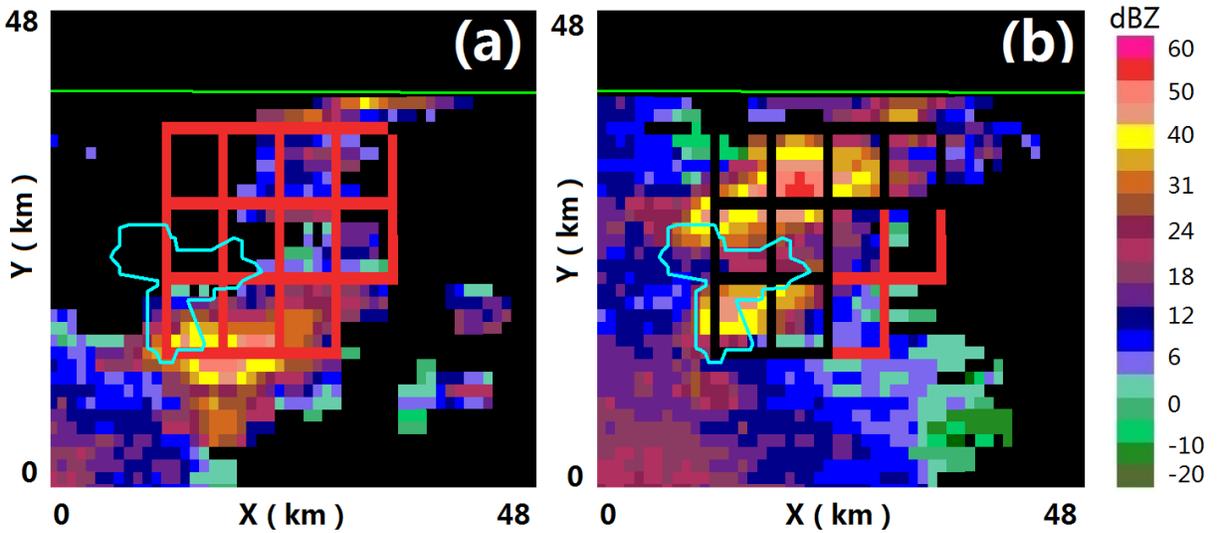

**Figure 7.** Same as Figure 6 but over a different sub-domain at the 30-min nowcast issue time, 2310 UTC 7 July 2012 (a) and at verification time, 2340 UTC (b) to show an example of CI nowcast.

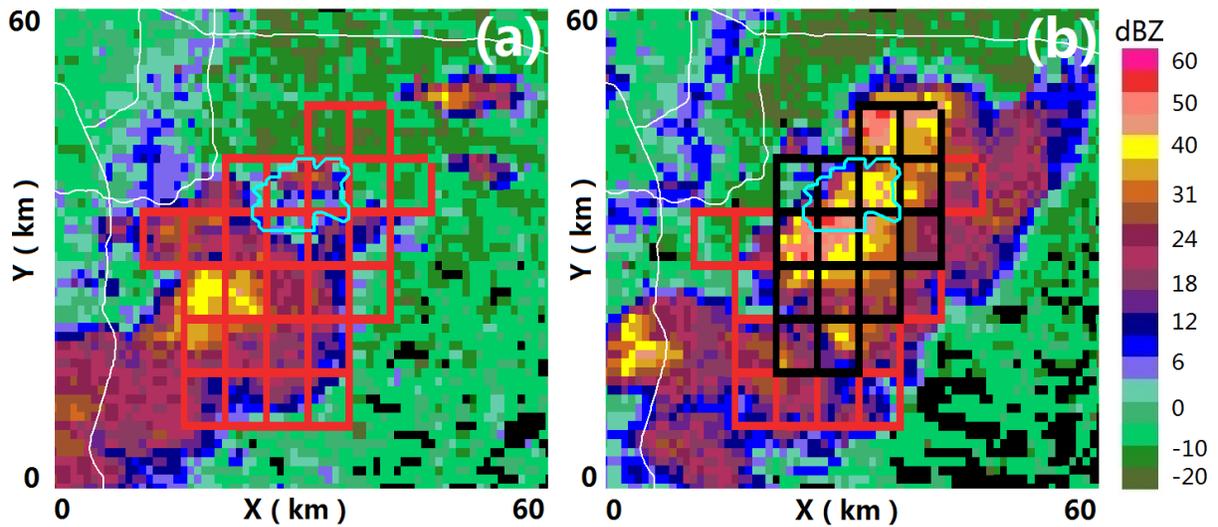

**Figure 8.** Same as Figure 6 but over a different sub-domain at the 30-min nowcast issue time, 2210 UTC 6 June 2012 (a) and at verification time, 2240 UTC (b) to show an example of CI nowcast.

SBOW is not only able to nowcast the storm growth and the convective initiation but also the storm propagation. Now we present the 30-minute forecast results for the propagation of the squall line case of 7 July 2012. Figure 9 shows the observed reflectivity of the squall line overlaid with SBOW and TITAN 30-min nowcasts. Comparing Figures 9 (a) and (d) at 22:55 UTC, we can see that the SBOW nowcasts can capture the squall line advection very well

(indicated by the white arrows). This is encouraging because the SBOW predicts the squall line movement without the need to calculate the computationally expensive motion vectors of the radar echoes as in the extrapolation techniques. The results at 23:10 and 23:25 UTC show that the SBOW nowcasts continue to capture the squall line movement quite well. As a comparison, TITAN also gives a good forecast for this case characterized mainly by linear propagation, but its forecast area is much smaller. As will be shown in next section, this will lead to a lower POD value.

From Figure 9, we can also identify several false alarms in the SBOW nowcast that occurred mainly behind the squall line, as indicated by the yellow arrows, although they decreased with time. These false alarms could be the result of the inadequacy of SBOW in predicting storm decay due to the limited number of predictors used in the current algorithm, which suggests that future improvement of SBOW should include the predictors representing storm decay. Opposite to SBOW, TITAN tends to under-predict the storm areas, resulting in fewer false alarms. Figure 10 shows similar results for the case of 6 June 2012.

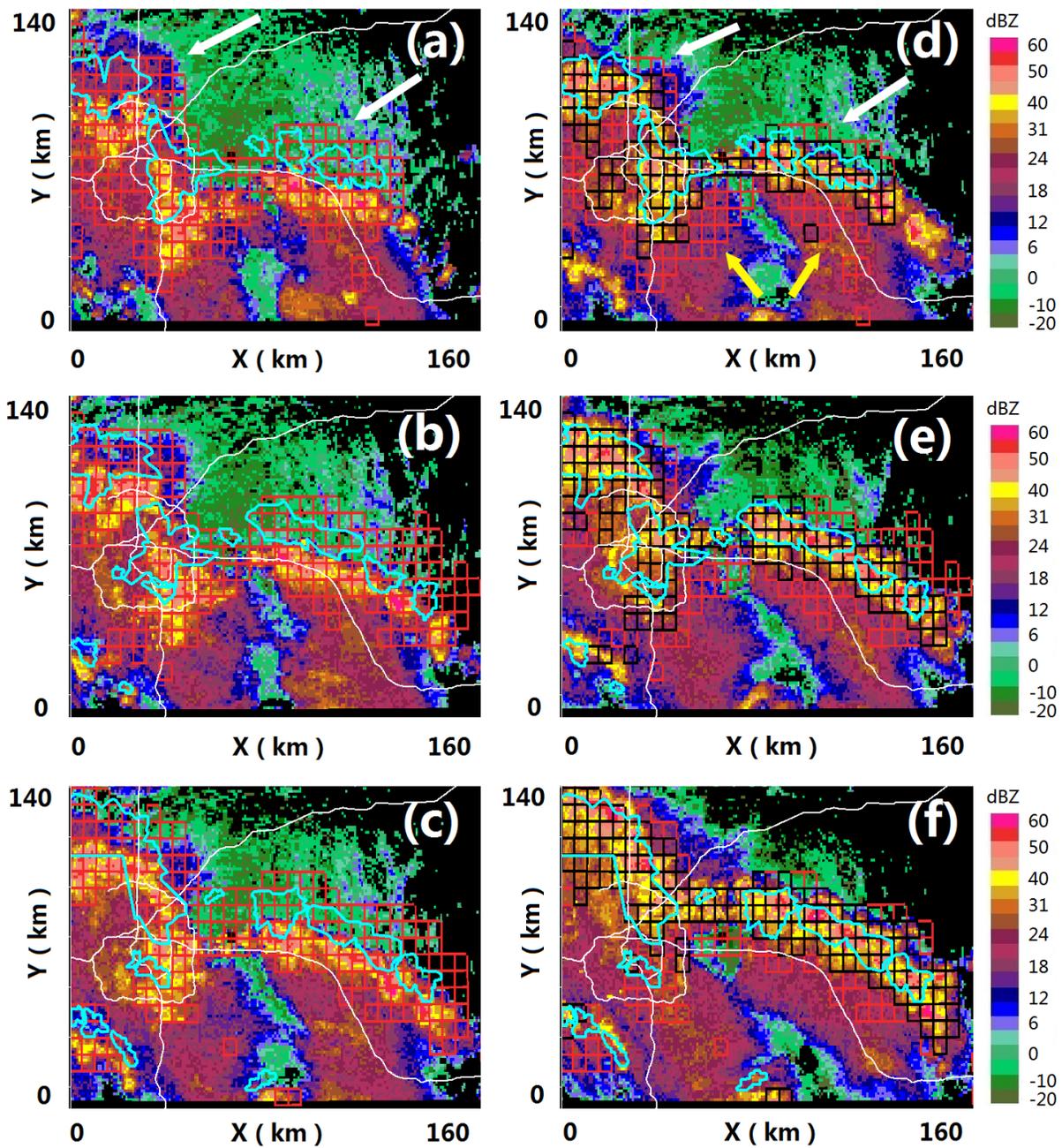

**Figure 9.** Same as Figure 6 but over a large area to show an example of storm propagation nowcast on 7 July 2012. (a), (b) and (c) are the SBOW and TITAN 30-min nowcasts at issue time, 2255, 2310 and 2325 UTC respectively. (d), (e) and (f) are these same nowcasts superpositioned over reflectivity at verification time 2325, 2340 and 2355 UTC respectively.

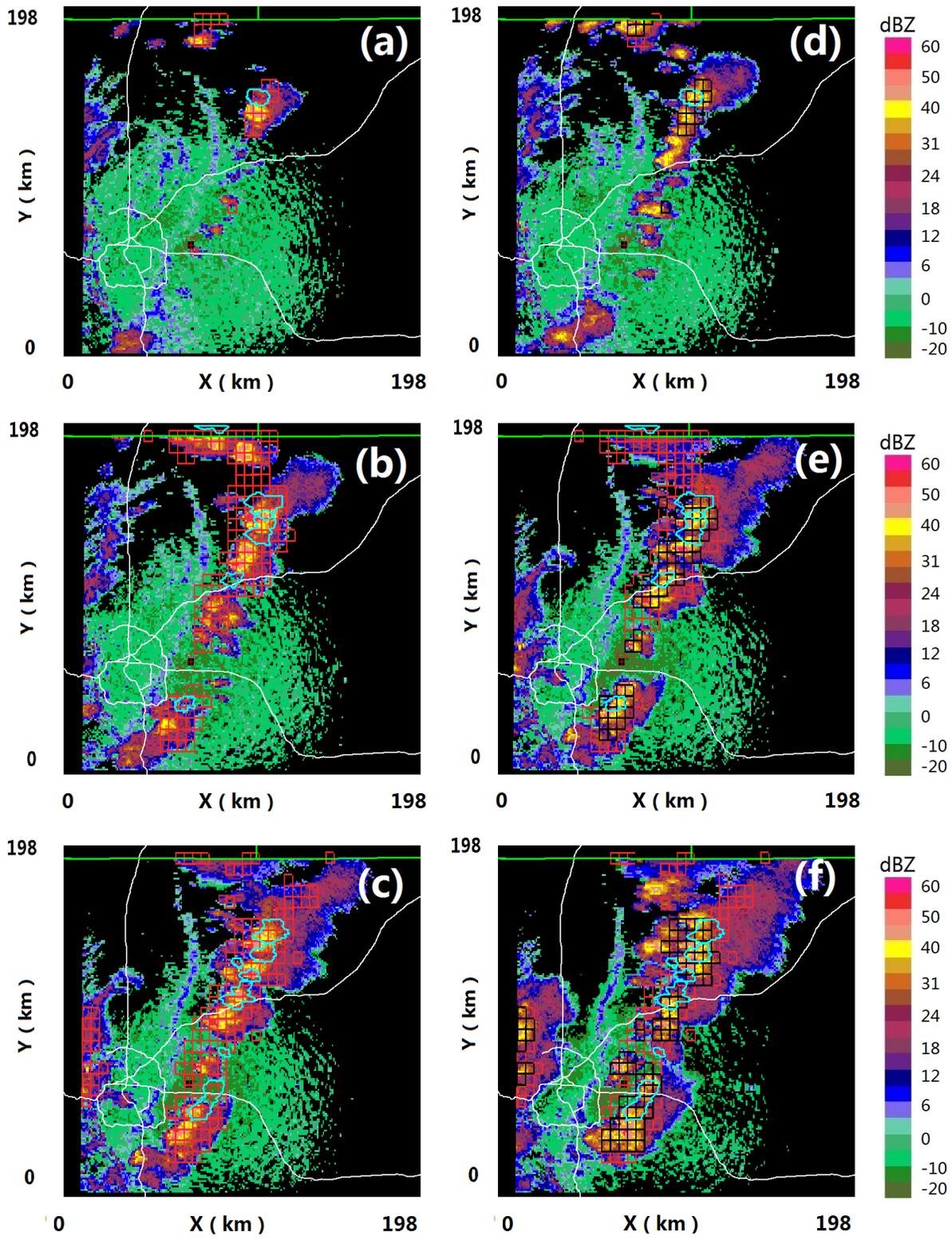

**Figure 10.** Same as Figure 6 but over a large area to show an example of storm propagation nowcast on 6 June 2012. (a), (b) and (c) are the SBOW and TITAN 30-min nowcasts at issue time, 2125, 2210 and 2240 UTC respectively. (d), (e) and (f) are these same nowcasts superpositioned over reflectivity at verification time 2155, 2240 and 2310 UTC respectively.

### 4.2.2 Quantitative Analysis

The overall POD, FAR, CSI and Heidke Skill Score (HSS; see *Wilks* 2011) values for the 30-min SBOW and TITAN nowcasts are given in Table 2 for the two studied cases. As the 7 July 2012 case is a squall line of linearly propagation, larger convective system, it is not surprising that it achieved higher CSI and HSS scores than that of the 6 June 2012 case. Table 2 shows that SBOW has substantial higher POD values than TITAN, but it also has higher FAR values, which leads to that both methods have very similar CSI values. The performance diagram (Fig. 11) shows more details of this. We can see that TITAN tends to have an underforecasting bias and SBOW tends to have a slight overforecasting bias, which means TITAN has more misses and SBOW has more false alarms.

With regard to HSS, TITAN has higher value than SBOW. Although TITAN only extrapolates existing storms, its forecast is still reasonable for large and stable systems. In this initial study, SBOW shows encouraging potential to nowcast CI or storm growth. But usually, the verification area of CI or storm growth is very small, which means these improvements do not impact the verification values significantly.

Table 2. Verification statistics for the SBOW and TITAN nowcast for the case 6 June and 7 July 2012. Note that the verification of TITAN nowcasts is also performed on the $0.06° \times 0.06°$ box.

| Date | POD | | FAR | | CSI | | HSS | |
|---|---|---|---|---|---|---|---|---|
| | SBOW | TITAN | SBOW | TITAN | SBOW | TITAN | SBOW | TITAN |
| 20120606 | 0.62 | 0.52 | 0.51 | 0.46 | 0.37 | 0.36 | 0.51 | 0.53 |
| 20120707 | 0.61 | 0.54 | 0.41 | 0.37 | 0.43 | 0.41 | 0.54 | 0.58 |

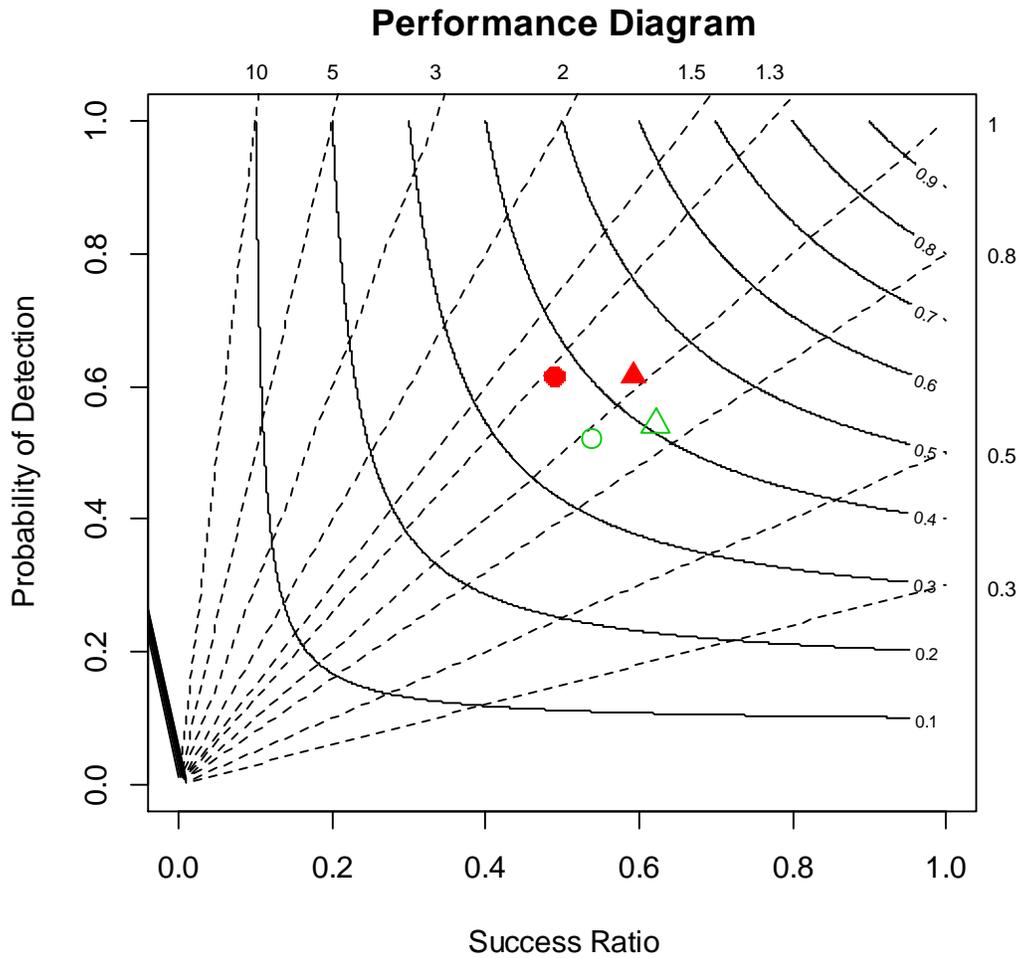

**Figure 11.** The performance diagram of SBOW and TITAN. Dashed lines represent bias scores with labels on the outward extension of the line, while labeled solid contours are CSI. Four results are shown: SBOW forecasts (bold red circle, 20120606; bold red triangle, 20120707) and TITAN forecasts (open red circle, 20120606; open red triangle, 20120707)

Our qualitative and quantitative evaluations showed encouraging results in terms of nowcasting convective initiation, growth, and propagation. Nevertheless, the results also suggest that problems exist, especially the problem of false alarms behind or at the location of old storms. Further development efforts are required to improve the performance of the SBOW method by choosing features that can predict storm decay. This will be the focus of one of our future research efforts

## 5 Summary and discussions

This study proposed a nowcasting method called SBOW under the machine learning framework using real-time VDRAS reanalysis data. SBOW divided the study domain into many position-fixed small boxes and attempted to answer the following nowcasting question: will a radar echo > 35 dBZ appear in a box in 30 minutes? Box-based temporal and spatial features, which include time trends and surrounding environmental information, are constructed. The machine learning framework is employed to perform the nowcasting task, i.e., use the box-based constructed features to train a SVM classifier, and then use the trained classifier to make predictions. The above designs of SBOW make the system small, compact, straightforward, and easy to maintain and expand at low cost. The only input data for SBOW are radar reflectivity and VDRAS analysis data. The experimental results showed that, although no complicated tracking algorithm was used, SBOW could predict storm movement and storm growth with reasonable skill. The strength of SBOW in comparison with the traditional extrapolation-based nowcast system TITAN is its ability in nowcasting the convective initiation and growth as demonstrated both in the qualitative verification and the statistically higher POD. SBOW can be expanded to use other model analysis data sources.

Although SBOW showed potential in predicting convective initiation and growth, its success is still limited. One reason for this is that the training data for CI cases are still insufficient, which means that the machine learning method cannot acquire enough knowledge to make correct decisions. As the duration of CI constitutes only a small proportion of the whole lifetime of a storm, and the storm area of CI is often very small, it is difficult to collect enough training data for CI cases. Another reason is that numerical models, such as VDRAS, still need improvement to obtain finer retrieval information. For cases of very small, isolated storm cells for which SBOW did not perform well, it is likely that resolutions higher than the current 3-km resolution VDRAS analysis will help.

False alarms often occur behind or at the location of old storms, e.g., in areas occupied by old storms that appear at forecast issue time but disappear at verification time. This is a difficult situation for SBOW in the current design due to the choice of predictors. Although the use of temporal trends (*dw* and *dpt*) could have played a role in predicting storm decay, this is not an adequate explanation. In future studies, we will test other predictors that may be linked to predicting storm decay. The possible candidates are downdraft, relative humidity, and maximum cooling.

Adding some other features to SBOW, such as humidity, Convective Available Potential Energy (CAPE), and Convective Inhibition (CIN) from VDRAS may also improve nowcast accuracy by better capturing a storm's environmental conditions. However, because these quantities are multi-scale in nature, it is not as straightforward to define as it was for the exact "features" of vertical velocity and perturbation temperature. Active research is being conducted to identify more relevant features that will lead to the improvement of the nowcasting scheme. However, if done incorrectly, adding more features could degrade the results [*Lee et al.*, 2004].

To improve SBOW, it is important to obtain more VDRAS data, from which SVM can acquire more knowledge to enable better decisions. This requires running VDRAS operationally on a fixed domain with fixed resolution and configurations. The Beijing Meteorological Bureau (BMB) has been running VDRAS over the last few years operationally, and our plan is to collaborate with BMB to train SBOW with more data. Our ultimate goal is to test our method operationally.

**Acknowledgments**

This work was supported jointly by China Special Fund for Meteorological Research in the Public Interest (grant GYHY201506004), the National Natural Science Foundation of China (grant 41405110) and Natural Science Foundation of Shandong Province (grant ZR2016DM05). All data from this study can be requested from the authors directly.